\documentclass[11pt]{article}
 
\usepackage{amsmath}
\usepackage{amssymb}
\usepackage{graphicx}
\usepackage{cite}
\usepackage{url}
\usepackage{setspace}

\usepackage{url}

\begin{document}

\title{Dispersal limitation and roughening of the ecological interface}

\maketitle

\noindent

\begin{center}
\author{ANDREW J. ALLSTADT,$^{1,\diamond,\star}$ JONATHAN A. NEWMAN,$^{2,\S}$ JONATHAN A. WALTER,$^{3,\sharp}$ G. KORNISS,$^{4,\ddag}$} and THOMAS CARACO$^{5,\dag}$\\
\end{center}

\noindent
1. \emph{Blandy Experimental Farm, University of Virginia, Boyce VA 22620, USA} \\ \\
\noindent
2. \emph{School of Environmental Sciences, University of Guelph, Guelph, Ontario N1G 2W1, Canada} \\ \\
\noindent
3.  \emph{Department of Environmental Sciences, University of Virginia, Charlottesville, VA 22904-4123, USA} \\ \\
\noindent
4. \emph{Department of Physics, Applied Physics, and Astronomy,\\
Rensselaer Polytechnic Institute, 110 8th Street, Troy NY 12180-3590, USA}\\ \\
\noindent
5. \emph{Department of Biological Sciences, University at Albany, Albany NY 12222, USA} \\ \\

$^{\diamond}$\emph{Present Address}:  Department of Forest and Wildlife Ecology, University of Wisconsin-Madison, Madison WI 53706, USA \\ \\

$^{\star}$Corresponding author, e-mail: allstadt@wisc.edu; $^{\S}$e-mail: jonathan.newman@uoguelph.ca; $^{\sharp}$e-mail: jaw3es@virginia.edu; $^{\ddag}$e-mail:korniss@rpi.edu; $^{\dag}$e-mail: tcaraco@albany.edu. \\ \\

\emph{Keywords}: ecological invasion, front-runner, interface dynamics, scaling laws, spatial competition, stochastic roughening \\ \\




\parindent=5mm

\newpage

\emph{Abstract}.
Limited dispersal distance, whether associated with
vegetative growth or localized reproduction, induces spatial
clustering and, in turn, focuses ecological interactions at the
neighborhood scale.  In particular, most invasive plants are clonal,
cluster through vegetative propagation, and compete locally.
Dispersal limitation implies that invasive spread occurs as advance
of an ecological interface between invader and resident species.
Interspecific competition along the interface produces random
variation in the extent of invasive growth. Development of these
random fluctuations, termed stochastic roughening, will often
structure the interface as a self-affine fractal; a series of
power-law scaling relationships follows as a result.  For a diverse
array of local growth processes exhibiting both forward and lateral
propagation, the extent of invader advance becomes spatially
correlated along the interface, and the width of the interface (the
area where invader and resident compete directly) increases as a
power function of time.  Once roughening equilibrates statistically,
interface width and the location of the most advanced invader (the
``front-runner'') beyond the mean incursion should both increase as
a power function of interface length.  To test these predictions, we
let white clover (\emph{Trifolium repens}) invade ryegrass
(\emph{Lolium perenne}) experimentally.  Spatial correlation
developed as anticipated, and both interface width and the
front-runner's lead scaled as a power law of length.  However, the
scaling exponents differed, likely a consequence of clover's growth
morphology. The theory of kinetic roughening offers a new framework
for understanding causes and consequences of spatial pattern in
between-species interaction, and indicates when interface measures
at a local scale predict properties of an invasive front at extended
spatial scales.


\vspace*{0.75truecm}
\begin{center}
{\large {I}}NTRODUCTION
\end{center}

Pattern analysis of plant communities commonly reveals spatial mosaics generated by clustered growth of individual species \cite{Cain_1995,Dale_1999,Condit_2000}.  Clustering may follow a template set by environmental heterogeneity, especially if different locations favor different species \cite{SnyderChesson_2003}, but more often, strong dispersal limitation aggregates conspecific individuals \cite{Harada_1994}.  For example, most invasive plants are clonal and propagate vegetatively \cite{Kolar_2001,Sakai_2001,Liu_2006}, so that invaders will initially cluster among residents \cite{Korniss_JTB2005,Cantor_2011}.

Spatial clustering influences frequencies of different biotic interactions and the consequent population dynamics \cite{Herben_2000}.  Individual plants usually compete at the nearest-neighbor scale \cite{Goldberg_1992,Levine_2004}.  Therefore, when different species each aggregate spatially and interact locally, intraspecific competition should predominate within clusters, while interspecific competition will localize at the interface between clusters \cite{Chesson_2000,Yurkonis_2004}.  The resulting interaction geometry implies that the advance \emph{versus} extinction of a rare competitor may depend on development and motion of a between-species interface \cite{Gandhi_1999,Allstadt_2009,OMalley_2010}. That is, we envision a non-equilibrium system where increase (decrease) in a species' abundance drives interface motion.  The dynamics of an ecological interface distinguishes it from an ecotone, when the latter implies a change in species
composition due to abiotic factors that vary slowly relative to the timescale of population growth \cite{Gastner_2009,Eppinga_2013}.

A focal species' density declines from positive equilibrium to rarity across the width of an ecological interface.  For clarity, we refer to the focal species as the invader, so that invasive movement of the interface (or advancing front) implies increase in the area occupied by the focal species.  Given this simple, general picture, we ask how varying the length of the interface affects statistical properties of an invader-resident interaction.  We emphasize the relative position of the ``front-runner,'' the furthest advanced invader, a metric used in both theoretical and applied invasion ecology \cite{Hajek_1996,Clark_2001,Thomson_2003}.

To begin, we assume invasive advance is a dispersal-limited, stochastic process and treat the ecological interface as a self-affine fractal, which implies that both interface width and the front-runner's lead will  depend on the length of the advancing front \cite{OMalley_2006,OMalley_BMAB}.  Second, we report an experiment testing our predictions at a local scale; we let Dutch white clover (\emph{Trifolium repens}) advance into plots of perennial ryegrass (\emph{Lolium perenne}).  Our contribution lies in the application of scaling laws, drawn from the theory of kinetic roughening \cite{Family_1985,Barabasi_1995}, as a framework for understanding spatial-growth patterns.  We help clarify how interface roughening organizes biotic interactions of locally clustered species.  More generally, the concepts we invoke may help integrate spatial-growth processes across quite different length scales.

\vspace*{0.75truecm}
\begin{center}
{\large {L}}OCAL {\large {D}}ISPERSAL and {\large {I}}NTERFACE {\large {R}}OUGHENING
\end{center}

Consider a spatially aggregated invader advancing into an area occupied by a resident species.  As the dispersal-limited invader's abundance increases from rarity, it displaces the resident along the interface.  Deterministic reaction-diffusion models, which assume continuous densities, cannot appreciate observable consequences of spatially correlated variability along a front generated by advance of a locally dispersing species \cite{Clark_2003,Antonovics_2006,OMalley_SPRINGER}.  But
discrete (``individual-based'') models may capture effects of the nonlinearity and stochasticity inherent to the dynamics of rarity at an ecological interface \cite{Durrett_1994tpb,Pachepsky_2011}.  Therefore, we characterize interface roughening, and motivate our experiment, in the context of a discrete, stochastic process where an invader and a resident compete for space \cite{Allstadt_2012}.  For commentary on discrete \emph{versus} continuous-density models, see Moro [2003] or van Saarloos [2003]; Appendix A summarizes differences between these models.

\begin{figure}[t]
  \centering
  \vspace*{3.00truecm}
  \includegraphics{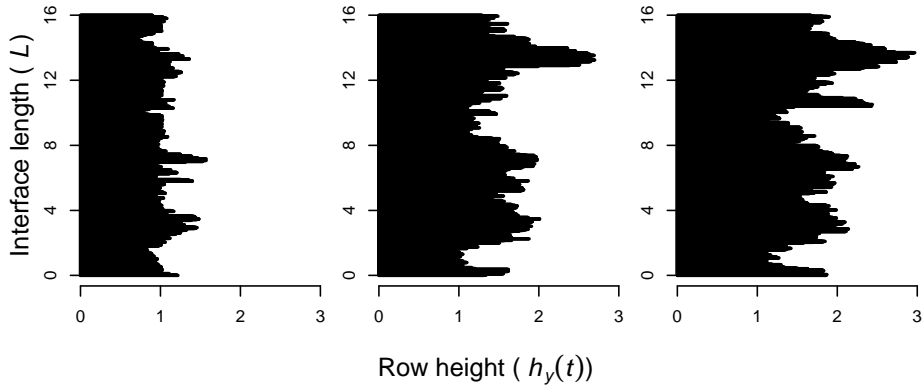}
  \vspace*{2.20truecm}
\caption{\small
    White clover (\emph{T. repens}, black area) advancing into perennial ryegrass (\emph{L. perenne}), from photographs taken during experiment.  Interface length $L$$\equiv$$L_y$$=$$16~m$.  June (left), August (center) and October (right) 2010 shown.  The interface advances, left to right, and roughens; neighboring heights suggest spatial correlation.
}
\label{PlotHeights}
\end{figure}

In a two-dimensional environment, development of spatially correlated growth along an interface is termed stochastic roughening \cite{Kardar_1986,Plischke_87,Barabasi_1995}. Roughening and dynamic scaling offer a conceptual framework for identifying dynamics shared by correlated growth processes differing in details of local interactions.  Applications span growth processes in physical materials \cite{Barabasi_1995}, biological tissues, including tumors \cite{Bru_2003,Galeano_EPL2003}, parallel-computing and information systems \cite{Korniss_2000,Korniss_2003}, and in ecological invasion \cite{OMalley_2006,OMalley_BMAB}.  When we analyze the front-runner's location, correlated fluctuations along the interface are particularly important, since traditional extreme-value statistics \cite{Fisher_1928,Galambos_1994}, developed for \emph{independent} random variables, do not apply \cite{Majumdar_2004}.  Figure \ref{PlotHeights} shows an advancing, roughening interface from the field experiment we report below. The invader's advance along the interface clearly roughens with time. The figure also suggests correlated advance at nearby locations.

\vspace*{0.75truecm}
\begin{center}
{\emph{Interface roughening: development and saturation}}
\end{center}

After defining attributes of the interface, we describe a roughened
front`s development.  Then we address roughening at statistical
equilibrium (termed ``saturation'').  Table 1 lists model symbols.

\begin{table}[t]
\centering
\label{definitions}
\begin{tabular}{|c|l|}
 \hline
Symbols & Definitions \\
 \hline
  $L_x ,\; L_y ( = L)$ & Lattice size ($L =$  interface length = front length) \\
  ${\rm \bf{r}}$ & Lattice site \\
  $t$ & time \\
  $h_y (t)$ & Rightmost invader in row $y$ at time $t$ \\
  $\overline{h}(t)$ & Mean of $h_y (t)$  (the average is taken across all rows $y$) \\
  $h_{max} (t)$ & Rightmost invader at time $t$ \\
  $\Delta_{max}(t)=h_{max}(t)-\overline{h}(t)$ & Distance from front-runner to mean of front \\
  $v^*$ & Asymptotic velocity of invasive advance \\
  $\langle w^2 \rangle$ & Mean squared interface width \\
  $\xi (t)$ & Correlation length along interface \\
  $t_{\times}$ & Crossover time, where $w^2$ equilibrates \\
  $\alpha$ & Roughness exponent \\
  $\beta$ & Growth exponent \\
  $z$ & Dynamic exponent \\
  \hline
\end{tabular}
\caption{Definitions of variables.}
\end{table}

An $L_x$$\times$$L_y$ rectangular lattice represents a habitat
occupied by the resident and invader species.  Each lattice site is
either occupied by the invader, occupied by the resident, or is
empty.  Mortality opens occupied sites.  An empty site $\bf{r}$
becomes occupied through propagation from an occupied site among the
nearest neighbors of $\bf{r}$.  Typical neighborhoods include only
the 4 or 8 closest sites; restricting propagation to nearest
neighbors, of course, imposes dispersal-limitation.  We assume that
the invader's competitive superiority drives interface motion.

Suppose that the invader initially occupies only a few vertical
columns at the left edge of the lattice, and the resident occupies
all other sites.  Invasive advance occurs in the $x$-direction.
Importantly, neighborhood geometry (the dispersal constraint)
permits both forward and lateral growth.  The former pushes the
front, and the latter generates spatial correlation along the front
\cite{Kardar_1986,Barabasi_1995}.  That is, lateral growth of
advanced heights tends to increase height in adjacent rows.

We let $L\equiv L_y$, interface length (hence, front length).  At
time $t$, $h_y(t)$ is the location of the most advanced (right-most)
invader in row $y$; $y = 1, 2, ..., L$.  The front's average
location is the mean height among rows, $\overline{h}(t) =
\sum_{y}h_y(t)/L$.  We take longitudinal system size $L_x$ as
sufficiently large that it does not affect population processes.

\begin{figure}[t]
  \centering
  \vspace*{3.00truecm}
  \includegraphics{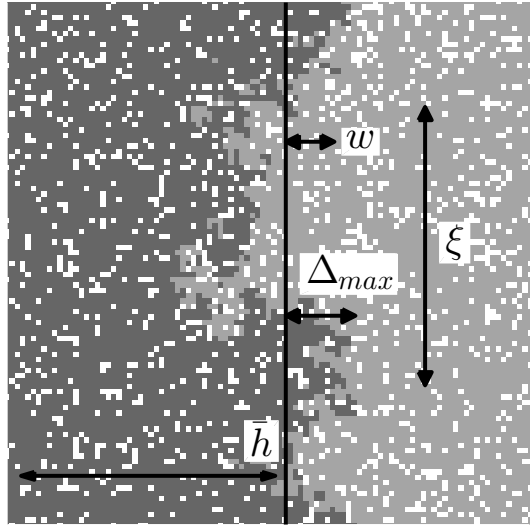}
  \vspace*{4.00truecm}
\caption{\small
Width ($w$) and the extreme advance
($\Delta_{max}$) relative to the mean front position
($\overline{h}$) in a rough front.
For illustration, correlation length $\xi$ is also indicated.  Dark: invader, medium: resident and white: open. Figure generated by simulation of model in O'Malley et al. (2006).}
\label{simulate}
\end{figure}

Figure \ref{simulate} shows the width of the interface about the
invader's average incursion $\overline{h}(t)$.  To quantify
roughening, we define the width of the interface \emph{via}:
\begin{equation}
w^2(L,t) = \frac{1}{L}\sum_{y = 1}^{L}
[h_y(t) -\overline{h}(t)]^2 \;
\label{w2}
\end{equation}
Roughness $w^2(L,t)$ itself varies stochastically, and we represent
its expectation (averaged over realizations of intrinsic noise) at
time $t$ by $\langle w^2(L,t)\rangle$.  We take $w=\sqrt{\langle
w^2(L,t)\rangle}$ as the width of the front, the typical extent of
the interface parallel to the direction of advance.

Next, we identify power-law scaling relationships that should
characterize the structure of an interface with spatially correlated
heights.  Importantly, these qualitative relationships do not, in
general, depend on details of the ecological interactions shaping
the interface.  Numerical calibration of the scaling laws can, of
course, differ across species and environments.

\emph{Interface development}. -- Assume a linear front at time $t = 0$, a flat initial interface.  As the invader begins to advance, the interface starts to roughen, and invader heights $h_y(t)$ become dependent random variables.  That is, a single correlation length $\xi (t)$ develops along the interface (Fig. \ref{simulate}).  Correlation length initially increases with time according to the power-law scaling $\xi(t) \sim t^{1/z}$ \cite{Majumdar_2005}, where $z$ is called the dynamic exponent.  But once $\xi (t)$ spans the length $L$ of the interface, ``crossover'' occurs.  The interface continues to advance, but roughening reaches statistical equilibrium when spatial correlation spans the length of the interface (roughening ``saturates'' at crossover) \cite{Barabasi_1995}.  The duration of interface development, termed crossover time $t_\times$, increases with interface length; the power-law scaling is $t_\times \sim L^z$.  The development of interface width offers a more easily tested prediction.  Prior to saturation, interface width $w$ exhibits temporal scaling behavior according to $w \sim t^{\beta}$.  $\beta$ $(\beta > 0)$ is called the growth exponent \cite{OMalley_2006}.

We monitor increasing correlation distance along the developing interface in two ways; each combines results from windows of length $l < L$.  The local width, $w_t(l)$, is the expected interface width estimated for the portion of the interface with length $l$, at time $t$. The height-difference correlation function \cite{Karabacak_2001} integrates roughness during both development and saturation.  The height-difference correlation function, at time $t$, is given by:
\begin{equation}
C_t(l) = \left\langle \left( h_{y+l}(t) - h_{y}(t) \right)^2 \right\rangle_{y}^{1/2}
\label{hhcorr}
\end{equation}
where $t>0$ and the average is taken across all rows $y$. Similarly, we define the height-height correlation function (the Pearson correlation):
\begin{equation}
G_t(l) = \frac{ \left\langle \left( h_{y+l}(t) - \overline{h}(t)\right) \left( h_{y}(t) - \overline{h}(t)\right) \right\rangle_{y}^{1/2}}
                     {\left\langle \left( h_{y}(t) - \overline{h}(t)\right)^{2} \right\rangle_{y} ^{1/2} } \;.
\label{Gcorr}
\end{equation}
\noindent
We use $G_t(l)$ to estimate correlation length $\xi(t)$; height-height correlation should decline as distance between rows increases.

For $l < \xi(t)$, both $w_t(l)$ and $C_t(l)$ exhibit power-law scaling over distances along the interface: $w_t(l),~C_t(l) \sim l^{\alpha}$.  $\alpha$ $(\alpha > 0)$ is the roughness exponent, and characterizes the fractal nature of the interface.  As the interface roughens with time, the correlation distance $\xi$ increases.  Consequently, the
linear dependence of $\ln w_t(l)$ and $\ln C_t(l)$ on $\ln l$, with slope $\alpha$, should extend to greater lengths $l$ along the interface, until saturation.

\emph{The saturated interface}. -- After crossover $(t > t_{\times})$, steady-state properties of the interface depend on its length $L$ \cite{Schehr_2006}.  Interface width $w$ scales with interface length according to
$\langle w^2(L,\infty)\rangle$$\sim$$L^{2\alpha}$.  That is, interface width increases as a power function of its length, according to the roughness exponent $\alpha$.

Note that we do not predict the value of roughness \emph{per se}, but ask how roughening changes from a shorter to a longer interface, or from portions of an interface to its entire length \cite{OMalley_BMAB}.  The power-law scaling for $\langle w^2(L,\infty) \rangle$ does not imply that clonal plants with a guerilla/runner morphology will be rougher at their growth interface than will clonal plants with a phalanx morphology.  Rather, the scaling relationship implies that if we can estimate interface roughening across local spatial scales, we can predict the way interface roughening increases at greater length scales.

In general, the scaling exponents are interdependent: $\alpha=\beta z$ \cite{Kardar_1986}; see Appendix B.  The dependence arises when the interface has a self-affine structure.  Geometrically, we begin with a one-dimensional interface embedded in a two-dimensional environment; random demographic events render the interface disorderly, and we assume that the interface equilibrates as an anisotropic fractal \cite{Barabasi_1995}.  ``Random demographic events'' means local processes that include both forward and lateral invader growth.  As noted above, forward growth pushes invasive advance, and lateral growth builds spatial correlations between heights.  Without lateral growth, each $h_y(t)$ becomes an independent birth-death process; independence implies, of course, lack of spatial correlation.  ``Anisotropic fractal'' implies a self-affine interface and an associated roughness exponent.  Consider the interface width $w(L)$.  Suppose that we increase interface length according to $L \rightarrow kL$.  Then the width must be re-scaled according to $w \rightarrow k^{\alpha} w$ to preserve statistical invariance (``look the same''); under anisotropic transformation, height and interface length must be increased by different factors.

\vspace*{0.75truecm}
\begin{center}
\emph{Roughening, scaling and the front-runner}
\end{center}

\emph{Interface velocity}. -- Analytic approximation of a discrete, stochastic model's asymptotic interface velocity $v^*$ remains a challenge \cite{Pechenik_1999}, especially for two-dimensional environments.  Dispersal limitation reduces velocity, compared to the corresponding reaction-diffusion model \cite{Moro_2001,Escudero_2004}.  Krug and Meakin (1990) found that a discrete model's \emph{reduction} in velocity, relative to the reaction-diffusion wave-speed, varies inversely with interface length.  Unfortunately, the exact reduction depends on the particular model's local dynamics.  That is, the manner in which velocity increases with $L$, as well as the asymptotic velocity $v^*$ itself, depend on details governing local (\emph{i.e}., individual-level) propagation and mortality.

Importantly, the scaling laws involving the roughness exponent $\alpha$ do not depend on invasion speed \cite{Barabasi_1995}.  This includes the probability distribution of the front-runner's lead.  However, this does not imply that velocity cannot be influenced by variation in roughness (see Discussion).

\emph{The front-runner: scaling of extremes}. -- The maximal invasive advance defines the front-runner's position.  At time $t$ we locate the front-runner at $h_{max}(t)=\max_{y}\{h_{y}(t)\}$.  Given mean interface height $\overline{h}(L,t)$, the invader's maximal relative advance at time $t$ is $\Delta_{max}(L,t) = h_{max}(t) - \overline{h}(L,t)$.  We assume that roughening equilibrates before considering the scaling of the expected lead $\langle\Delta_{max}\rangle_L$; note dependence on interface length $L$.

The probability density of the front-runner's excess $\Delta_{max}(L,t)$ has been obtained analytically \cite{Majumdar_2004,Majumdar_2005}.  For broad classes of dispersal-limited stochastic growth models, the scaled variable $\Delta_{max}/\langle\Delta_{max}\rangle$ has an Airy probability density, and the steady-state average excess of the front-runner over the mean height scales with interface length exactly as does the width.  That is, $\langle\Delta_{max}\rangle_L \sim L^{\alpha}$ \cite{OMalley_BMAB}.  Furthermore, we can infer the size of the extremes for an interface of linear size $L$ with estimates obtained in limited observation windows with size $L_{\rm {obs}}$.  We have: $\langle\Delta_{max}(L)\rangle \approx \langle \Delta_{max}(L_{\rm {obs}})\rangle k^{\alpha}$, where $k =  (L/L_{\rm {obs}})$, by the properties of a self-affine interface.  Table 2 collects scaling relationships we study; for a graphical summary of interface-roughness scaling, see Appendix B.

\begin{table}[t]
\centering
\begin{tabular}{|l|c|l|}
 \hline
Regime & Prediction & Comment \\
 \hline
  \emph{Development}  & $\xi(t) \sim t^{1/z}$ & Correlation length, dynamic exponent \\
  ~ & $w_t \sim t^{\beta}$ & Interface width, growth exponent \\
  ~ & $t_{\times} \sim L^z $ & Crossover time, interface length \\
  ~ & $C_t(l) \sim l^{\alpha}$ & Height-difference correlation, $l < \xi(t)$ \\
  \hline
  \emph{Stationarity} & $w \sim L^{\alpha}$ & Interface width, roughness exponent \\
  ~ & $\langle\Delta_{max}\rangle_L \sim L^{\alpha}$ & Front-runner's lead \\
  ~ & $\beta = \alpha / z $ & Self-affine fractal \\
  \hline
\end{tabular}
\caption{Predicted scaling relationships.  The interface roughens during development.  After spatial correlation spans interface length, interface width remains statistically stationary.}
\label{scalinglaws}
\end{table}

\vspace*{0.75truecm}
\begin{center}
{\large {A}}N {\large {E}}XPERIMENTAL {\large {I}}NTERFACE
\end{center}

To evaluate interface scaling, we studied dispersal-limited
competition between Dutch white clover (\emph{T. repens}) and
perennial ryegrass (\emph{L. perenne}).  Both species reproduce
mainly through local, clonal growth
\cite{TurkingtontEtal_79,Schwinning_1996a}.  \emph{T. repens}
propagates vegetatively through stoloniferous stems
\cite{Fraser_1989}, while \emph{L. perenne} produces tillers
\cite{Fustec_2005}.  Biotic interactions, including competition,
between these important forage crops are well understood
\cite{Cain_1995,Schwinning_1996b}.  We located experimental plots at
the University of Guelph Turfgrass Institute in an area homogeneous
with respect to micro-topography ($43^{\circ}33' N, 80^{\circ}13'
W$).  To minimize spatial heterogeneity, vegetation and top layer of
soil were removed, and the soil tilled before the experiment began.

\vspace*{0.75truecm}
\begin{center}
\emph{Experimental design}
\end{center}

We established plots with interface length $L =$ 1, 2, 4, 8, and 16
$m$, with four replicates of each length.  To avert edge effects, we
added a $0.5~m$ buffer, where no data were collected, at both ends
of every plot.  Each plot had a total height of $3~m$, and was
initially split lengthwise by plastic dividers into sections of
$1~m$ and the remaining $2~m$.  We planted \emph{T. repens} in the
one-meter sections, and \emph{L. perenne} in the two meter sections;
we anticipated that clover would advance, given the soil resources
and occasional mowing.  Appendix C details experimental methods.

By spring 2009 mono-cultures were well established, and we removed
the plastic barriers between species.  In June 2010 we began
recording the advance of \emph{T. repens} in each plot monthly.  We
resolved measurements at a scale of $1~cm^2$, the approximate size
of an individual clover ramet \cite{Silvertown_1981}.  We marked
each $1~m^2$ subsection of every plot permanently, to reference
growth measurements.  Each such subsection was photographed from
above after mowing each month. We re-projected each photo to correct
for perspective, and combined photos from the same plot.  We
recorded row heights $h_y(t)$ for \emph{T. repens} in each plot, and
noted the front-runner's lead on the mean clover height.

We tested power-law relationships against alternative linear and
exponential models \cite{Solow_2003}.  Additionally, we fit the
power-law models with two different assumptions regarding error
distribution.  The first assumed normally distributed, additive
error; the second assumed log-normally distributed, multiplicative
error \cite{Xiao_2011}.  Our scaling laws, Table \ref{scalinglaws},
predict the latter form.  We compared relative support for each
model using differences in AIC scores ($\Delta AIC$); we considered
models with $\Delta AIC < 2$ as supported substantially
\cite{Burnham_2002}.

\vspace*{0.75truecm}
\begin{center}
\emph{Results}
\end{center}

During the 2010 growing season, clover advanced rapidly; several
longer fronts approached the far end of the plot by October.  Figure
3A shows each plot's mean height $\overline{h}(t)$ against time.
Overall mean clover height increased for five consecutive months.
However, several clover fronts began to experience winter die-back
in October.  Therefore, our analysis treated data from June through
August as the interface-development period, and treated data from
September (month 4) as stationary.  This is an approximation, since
correlation lengths for larger values of $L$ continued to grow
during October.

\begin{figure}[t]
  \centering
  \vspace*{3.00truecm}
  \includegraphics{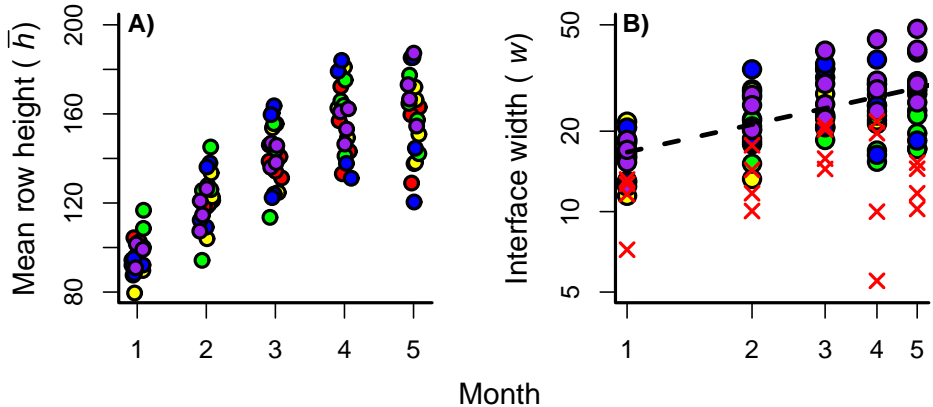}
  \vspace*{2.50truecm}
\caption{\small
Experimental interface development.  A. Mean plot heights ($cm$) by month.  Red, yellow, green, blue, and purple indicate, respectively, $L = 1, 2, 4, 8, 16 ~m$.  ``Noise'' added to abscissa for visibility.  B.  Each plot's interface width by month; note the logarithmic scale of the abscissa.  Dashed line indicates scaling of the first 4 months (development).  Estimated growth exponent $\beta = 0.31$.  $1~m$ plots marked as X, signifying earlier saturation; see text.}
\label{HeightWidthMonth}
\end{figure}

\emph{Interface development}. -- Spatial correlations between row heights $h_y(t)$ both increased in strength and extended to greater distances along the interface as clover advanced.  Since development of correlation length should not depend on $L$, we pooled observations from all plots.  We estimated correlation $G_t(l)$ between row heights $h_y(t)$, as a function of distance, for each of the five months.  Figure 4A shows the resulting correlogram; spatial correlation increased every month across most distances less than 200 $cm$.

The height-difference correlation $C_t (l)$ also revealed that correlation distance $\xi(t)$ increased during interface development.  More importantly, the distance $l$ ($l < \xi(t)$) over which $C_t (l)$ scaled as a power-law increased each month; see Figure 4B.  The scaling of the height-difference correlation depends on the roughness exponent $\alpha$, since $C \sim l^{\alpha}$ for $l < \xi(t)$.  Using the result for month 4, our model selection procedure strongly supported a power-law relationship with multiplicative error (Table \ref{AICscores}).  Regression analysis of the $C_t(l)$ results yielded an estimate of the roughening exponent as $\alpha = 0.277 \pm 0.002$; see Figure 4B.

During interface development, the temporal increase in roughening should scale such that $\langle w^2(L,t)\rangle$$\sim$$t^{2\beta}$.  Figure 3B shows each plot's interface width against time.  We first tested the predicted scaling after excluding data from plots with $L=1~m$, since roughening in those plots equilibrated earlier than was the case for larger $L$.  Our model selection procedure found support for the power-law model with multiplicative log-normal error (Table 3). Using this model, we estimated the growth exponent $\beta$  as 0.34 $\pm$ 0.12 (mean $\pm$ 95\% confidence interval; $R^2 = 0.355$).  Including the plots where $L=1~m$ had little effect, producing an estimate $\beta$ = 0.313; see Figure 3B.  In parallel, Figure 4C shows how scaling of the local interface widths $w_t(l)$ developed through time.

\emph{Velocity}. -- After saturation, we anticipated that velocity would increase with interface length.  August-to-September velocities (differences in mean monthly clover heights) were all positive, but independent of interface length $L$. Interestingly, the greatest overall mean velocity occurred  during the first month of growth.  As the growing season ended in September, longer fronts continued to advance, but some shorter fronts receded.

\begin{figure}[t]
  \centering
  \vspace*{2.50truecm}
  \includegraphics{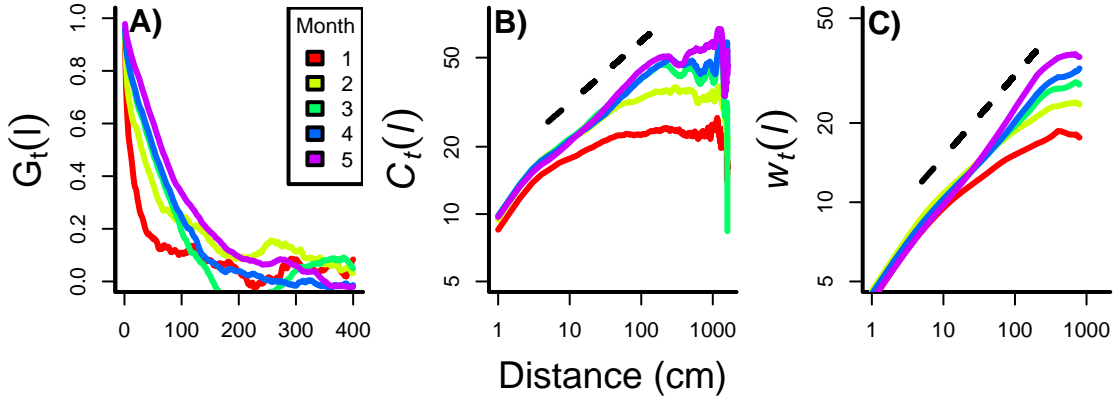}
  \vspace*{3.00truecm}
\caption{\small
Field experiment: interface development.  A.  Spatial correlogram: correlation of row heights, $G_t(l)$ [Eq. (3)].  The strength and lag distance at which correlations $G_t(l)$ remained significant increased through time, indicating an increase in the correlation length, $\xi(t)$, along the interface.  Key indicates month 1 through 5 for each plot.  B.  Height-difference correlation function, $C_t(l)$ [Eq. (2)], for months 1 through 5.  Distance over which power-law scaling holds increases with time, that is, increases as correlation distance increases.  Dashed line indicates month-4 scaling: $\alpha = 0.277$.  C.  Local interface width $w_t(l)$ across months.  The dashed line indicates the scaling relationship for month 4, based on the estimated growth exponent ($\beta$) of 0.34.
}
\label{SpatialCorr}
\end{figure}

\emph{Stationary roughness and the front-runner: power laws}. --  We
assume that roughening equilibrated in month 4.  We tested the
predicted scaling against alternative models in two ways.  The first
invokes the local roughening analysis, restricted to the final
month's data.  The second estimates how  mean interface width
increases with $L$.

After saturation, the local width $w(l)$, where $(l \leq L)$, should
scale as $w(l) \sim l^{\alpha}$.  We combined month-4 data from
different plots to characterize local roughening; see Figure 4C.
Our AIC-criterion strongly supported the power-law formulation with
multiplicative error (Table \ref{AICscores}).  The associated
estimate of the roughness exponent was $\alpha = 0.311 \pm 0.002$.
Our mean roughening analysis treated each plot's width $w(L)$
separately.  Using estimates from September (see Figure 5A), the
model selection procedure again provided substantial support for a
power-law relationship with multiplicative error (Table
\ref{AICscores}).  The power-law model for mean roughening as a
function of interface length $L$ led us to estimate $\alpha$ as
$0.278 \pm 0.18$.

Once roughening has equilibrated, the average lead of the
front-runner, beyond the mean interface height, should scale with
length as $\langle\Delta_{max}\rangle_L \sim L^{\alpha}$.  Our model
selection procedure once again found support for power-law scaling
with multiplicative error (Table \ref{AICscores}), the form
predicted.  Using the preferred model, we estimated the roughening
exponent as $\alpha = 0.475 \pm 0.19$; $R^2 = 0.6$; see Fig. 5B.

\begin{figure}[t]
\centering
\vspace*{2.50truecm}
       \includegraphics{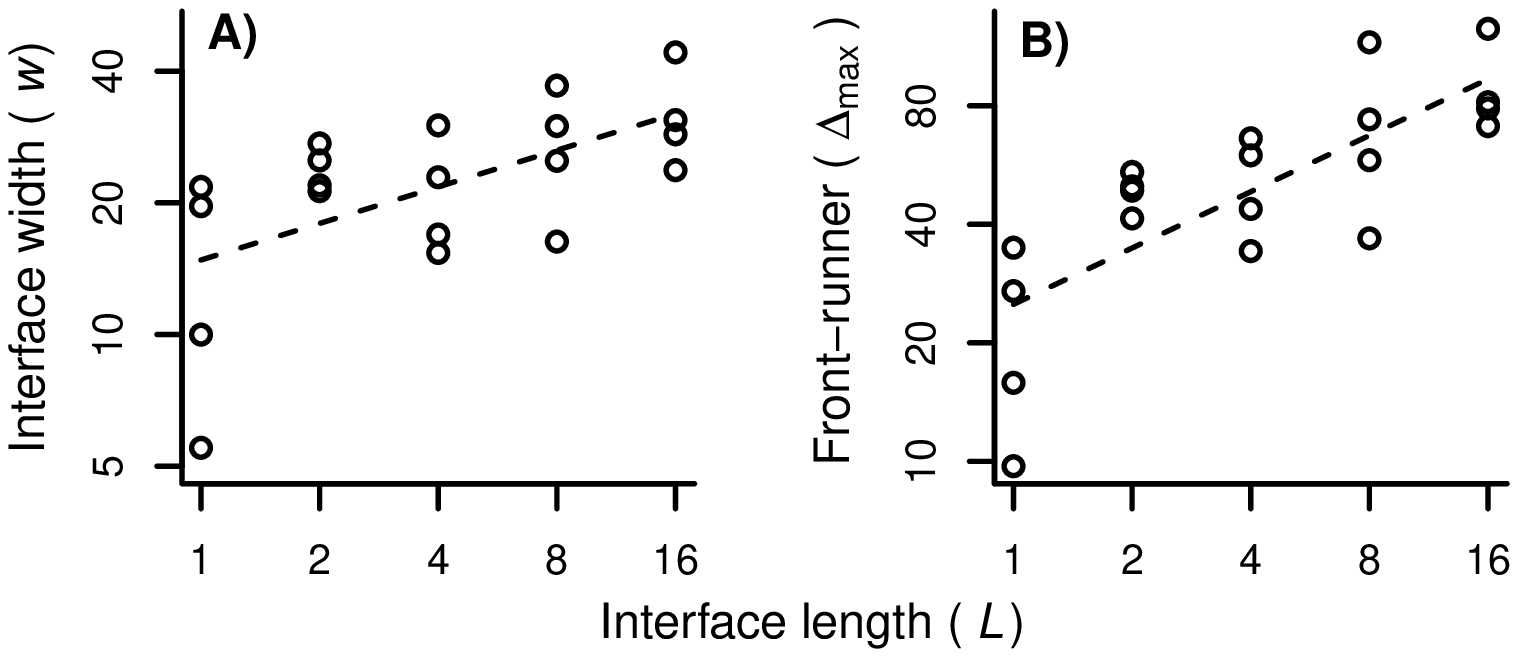}
\vspace*{3.00truecm}
\caption{\small
       Saturated roughening and the front-runner.  A.  Interface widths ($cm$)for different front lengths $L$, data from September (month 4).  Dashed line indicates power-law scaling.  B.  Front-runner's exceedance $(\Delta_{max})$, in $cm$, for different interface lengths $L$, data from month 4.  Dashed line indicates power-law scaling.
       }
\label{DeltaOfL}
\end{figure}
\begin{table}[h]
\centering
\begin{tabular}{|l|r|r|r|r|r|}
  \hline
  Clover Growth Analysis & Linear & Exp 1 & Exp 2 & Pow 1 & Pow 2 \\ \hline
  Dynamic $w^2(t)$ & 411.21 & 407.82 & 4.65 & 0 & 408.72 \\ \cline{2-6}
  $C(l)$ & 2014.94 & 1716.08 & 754.1 & 0 & 1398.53 \\ \cline{2-6}
  $w^2(l)$ & 3629.0 & 3061.25 & 1528.87 & 0 & 2528.99 \\ \cline{2-6}
  $w^2(L)$ & 118.44 & 117.02 & 2.56 & 0 & 117.18 \\ \cline{2-6}
  $\langle\Delta_{max}\rangle_L$ & 158.15 & 155.15 & 6.48 & 0 & 155.76 \\ \hline
\end{tabular}
\caption{$\Delta AIC$ scores.  Models compared are as follows.  Linear: $y = x + \epsilon$; Exp 1: $y = log(x) + \epsilon$; Exp 2: $log(y) = x + \epsilon$; Pow 1: $log(y) = log(a) + b ~log(x) + \epsilon$; Pow 2: $y = a x^b + \epsilon$.  $\epsilon$ is a random error term with zero expectation and finite variance.  Results support power-law models of Table 2.
}
\label{AICscores}
\end{table}

\vspace*{0.75truecm}
\begin{center}
{\large {D}}ISCUSSION
\end{center}

Clonal organisms dominate many communities
\cite{Gough_2002,Kui_2013}, so that dispersal limitation must
commonly generate strongly clustered growth patterns.  Within such
communities, invasive growth and competitive interactions will occur
within the width of an advancing interface, where invader and
resident neighborhoods overlap.  This general depiction of spatial
competition, common to numerous detailed models, invites application
of insights from the theory of kinetic roughening as a way to
understanding development, structure and motion of an ecological
interface.

Our scaling model is based on combined analytical and computational
study of stochastic partial differential equations for surface
growth.  A lattice-based model should, for proper choice of length
scale, induce a continuum equation which approximates an interface
defined by discrete heights $h_y(t)$ with a smooth curve
\cite{Barabasi_1995}.  The resulting equation for $\partial
h_y(t)/\partial t$ can include both growth terms depending on the
$\partial h_y(t)/\partial y$, the local gradient in height, and a
noise term.  Scaling relationships suggested by analysis of the
continuum equation can be verified in simulation \cite{Kardar_1986}.

In our field experiment, clover advanced, displacing ryegrass, in
every plot.  As the clover advanced, spatial correlation length
along the interface increased, accompanied by an increase in
roughness that scaled as a power function of time.  Interface
velocity did not depend on length; seasonal effects governing growth
and die-back were influential.

Each statistical analysis involving either the growth or the
roughness exponent supported a power-law formulation over logically
alternative linear and exponential models.  For the growth exponent,
we estimated  $\beta$ close to $0.34$.  In the stationary regime,
roughening increased as a power function of interface length, as did
the front-runner's lead.  Using roughening statistics, estimates of
$\alpha$ were close to $0.3$.  The front-runner's lead showed
power-law scaling, and we estimated $\alpha = 0.48$.  Confidence
intervals were, in some cases, relatively large.

Our understanding of interface roughening at the level of clustered
individuals should hold at extended scales.  For example, expansion
or contraction of a species' geographic range also involves
interface movement, whenever the ecological processes generating the
change are restricted to the proximity of the range boundary.

\vspace*{0.75truecm}
\begin{center}
\emph{Implications}
\end{center}

Power-law scaling of interface structure suggests that observed
statistical patterns, though not necessarily the underlying
processes, have no characteristic length scale.  If this were true,
we could predict interface behavior across scales.  For example,
every 10-fold increase in the length of a clover interface would
increase the front-runner's expected lead by a multiplicative factor
of $\sqrt{10}$.  An alternative assumes that spatial heterogeneity
in demographic rates, varying at a scale much greater than local
dispersal distance, implies that the roughness exponent $\alpha$
will vary along an extended interface.  In this case the front has a
multi-affine, sometimes called turbulent, structure, and local
estimates will not predict larger-scale behavior
\cite{Barabasi_1995}.  More generally, spatial heterogeneity,
whether fixed or temporally variable, can affect the likelihood an
invasion begins \cite{Duryea_1999,OMalley_2010} and front velocity
when invasion succeeds \cite{Shigesada_1986}.

Ordinarily, spatially structured interactions extend the inherent
time scale of population dynamics, compared to a homogeneously mixed
system \cite{OMalley_2010}.  At local scales, interface roughening
increases the number of interactions between the advancing invader
and the resident species, per unit interface length.  During
interface development, the rectangular area within which invader and
resident individuals interact (neighborhoods intersect) grows larger
with time; $L w(t) \sim L t^{\beta}$.  After roughening saturates,
we have $L w \sim L^{\alpha + 1}$.  A rougher interface (increased
$\alpha$) might then accelerate local population dynamics by mixing
clustered species spatially, and so increasing the density of
competitively asymmetric, neighborhood-scale interactions.

Our quantification of an ecological interface, both theoretically
and experimentally, used the heights $h_y(t)$, the most advanced
invader in each row $y$. We ignored structural lacunae due to
"overhangs" observable within the simulated interface shown in
Figure 2.  This simplification lets us apply scaling properties of a
fractal surface, without disguising the increased frequency of
between-species contact along the roughened interface.

\begin{figure}[t]
  \centering
  \vspace*{4.00truecm}
  \includegraphics{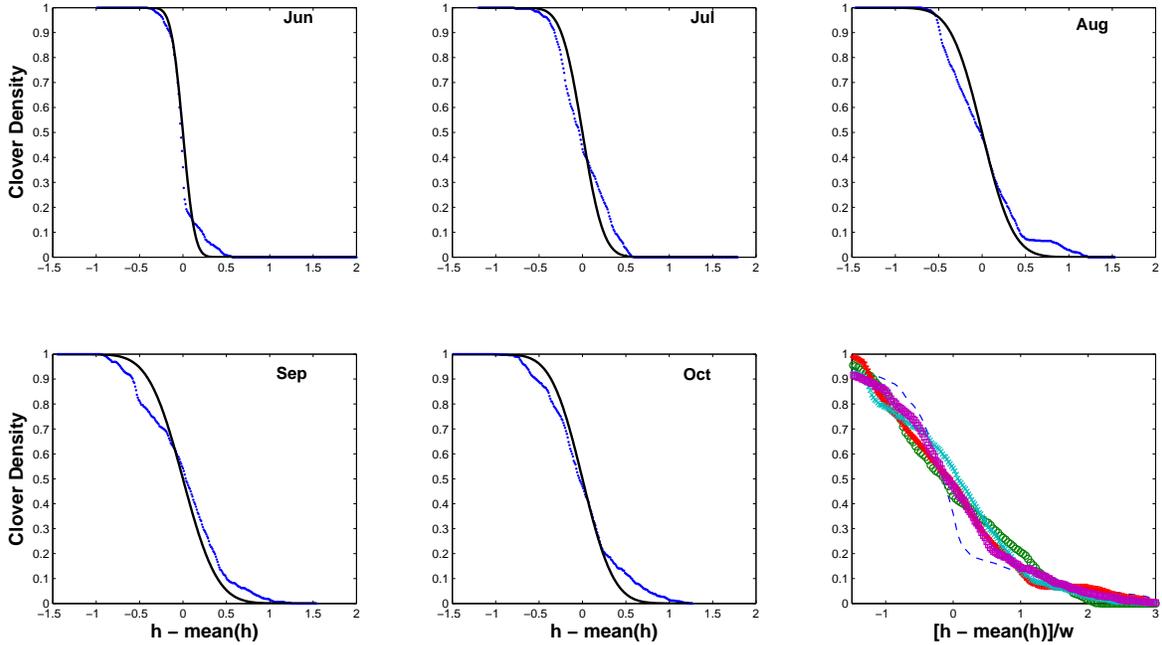}
  \vspace*{5.20truecm}
\caption{\small
    Density profiles of interface width, June through October (as indicated).  $L = 16~m$.  Each month's empirical profile indicated by $(\bullet)$.  Associated complementary error function for each month, parameterized by observed interface width, approximates data.  Interface widths for consecutive month are, respectively, $w = 0.15, 0.27, 0.4, 0.44, 0.4$.  Lower right. Data collapse.  Dividing height (relative to front's mean position) by interface width indicates that last four months' clover-density profiles share structural organization.  First month (broken line) insufficiently roughened to ``fit.'' Symbols are empty circle (Jul), square (Aug), $\times$ (Sep) and closed circle (Oct).
}
\label{ProfileOfWidth}
\end{figure}

Figure \ref{ProfileOfWidth} shows interface profiles from one experimental plot ($16~m$, same as Fig. 1) for all five months.  Each profile plots the fraction of $L$ rows occupied by clover as a function of distance from that month's mean height.  Every month clover and ryegrass occurred with nearly equal frequency at the mean height.  The first month's (June) profile drops sharply; the competitors mix very little as the interface begins to develop.  Closer to saturation, the profiles are less steep as interface width increases.

We can approximate observed interface profiles with the complementary error function.  Let $\rho_t (h)$ represent clover density at height $h$ and time $t$.  Then:
\begin{equation}
\rho_t(h) = \frac{1}{2} ~erfc \left([h - \overline{h}(t)])/w_t\right)
\label{rho}
\end{equation}
where $w_t$ is interface width at time $t$.  The complementary error function is:
\begin{center}
\begin{displaymath}
erfc~(x) = \frac{2}{\sqrt{\pi}} \int_x^\infty exp\left[ - z^2\right] dz
\end{displaymath}
\end{center}

\noindent
Equation \ref{rho} reasonably approximates observed profiles.  Mean invader density has an approximately Gaussian decline across the interface; see Foltin et al [1994].

\begin{figure}[t]
  \centering
  \vspace*{3.30truecm}
  \includegraphics{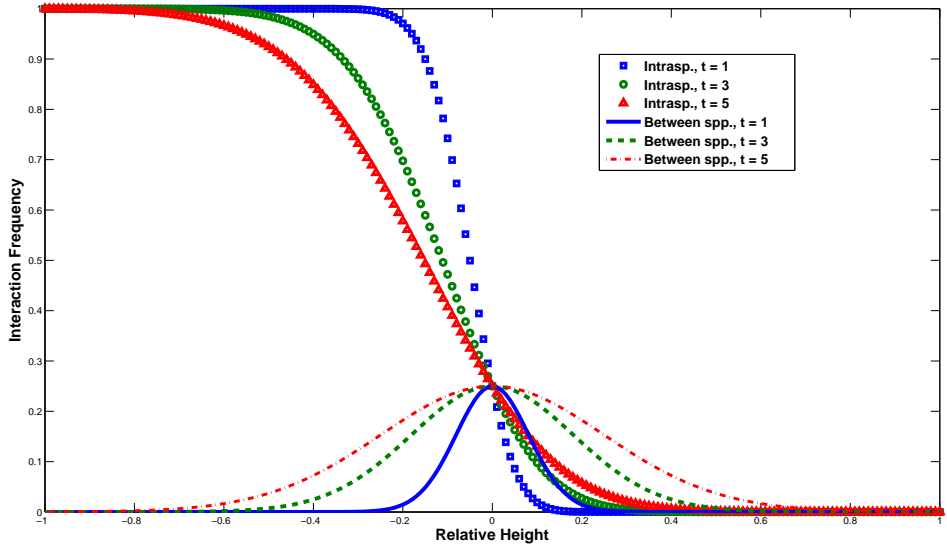}
  \vspace*{4.5truecm}
\caption{\small{Invader interaction frequencies.  $w(t = 1) = 0.13$ (blue), $w(t = 3) = 0.29$ (green), and $w(t = 5) = 0.39$ (red).  As time ($t$, in months) advances, interface width increases. As width increases, (1) decline in intraspecific competition behind mean height exceeds increase in intraspecific competition in front of mean height, and (2) interspecific competition increases symmetrically about mean height.}
}
\label{competition}
\end{figure}

The final subplot in Fig. \ref{ProfileOfWidth} (lower right) indicates ``data collapse'' of the last four months' density profiles.  Re-scaling height as $[h - \overline{h}(t)])/w_t$ reveals that the profiles share a common structural dependence on interface width, close to/at saturation.  That is, the re-scaled plot shows the basic relationship for which the July through October profiles are examples.  The first profile, a relatively un-roughened interface, has a different dependence.
Interface profiles average across spatially correlated invader density, but indicate how biotic interactions are organized within the interface.  Here, we neglect the density of open sites (which follows reasonably from our field results).  If intraspecific interactions occur in proportion to $\left[ \rho_t(h)\right] ^2$, their frequency will decline faster than invader frequency within the interface.  Interspecific interactions, if proportional to $\rho_t(h) [1 - \rho_t(h)]$, will increase initially, peak at $\overline{h}(t)$, and then decline.  Given this approximate picture, greater roughening (hence, larger width) increases interspecific mixing at the interface in a quantifiable manner; see Figure \ref{competition}.

We assumed that the interface developed from an initially linear array of individuals.  But invader growth will often commence as small, nearly circular clusters of individuals.  Many of the smallest clusters can disappear due to demographic stochasticity, despite an invader's ecological superiority \cite{Korniss_JTB2005}; large clusters will continue to grow.  At the critical cluster size \cite{Allstadt_2007} decline and growth are equally probable.  After a cluster attains sufficient size, we can treat its perimeter as a roughened ``surface;'' we discuss these issues in detail elsewhere \cite{OMalley_BMAB}.

Our general model assumes that an invader will propagate both forward and laterally; any unoccupied, nearest-neighboring site can be colonized at the same stochastic rate.  Cain et al. [1995] carefully mapped the architecture of clonal growth in a white clover population.  Node-to-node branching angles of apical meristems centered on $0^{\circ}$ (straight ahead), but some large angles were observed.  Lateral meristems branched off with a bimodal angular distribution, concentrated at $\pm 60 - 70 ^{\circ}$.  Clover, then, exhibits both forward and lateral growth, but with an overall bias toward forward propagation.  The resulting morphology could have induced the difference we observed between the scaling of the front-runner's lead and roughening with the length of the interface.

Schwinning and Parsons [1996a,b] modeled legume-grass spatial systems, and identified conditions where a legume (\emph{e.g}., clover) excludes grass competitively, where grass excludes the legume, and where the two coexist; also see Cain et al. [1995]. Low soil nitrogen (N) favors the legume; N-fixation generates a competitive advantage.  Defoliation/herbivory, as opposed to its absence, can promote the legume's competitive superiority at certain N-concentrations.  High soil-N favors grass over the legume, since grass then has the greater relative growth rate \cite{Schwinning_1996b}.  Intermediate conditions can maintain coexistence.  In our field experiment, we applied just enough N-fertilizer to insure that the grass would grow sufficiently to define a clear invader-resident interface.  We mowed plots periodically to prevent grass from growing tall enough to shade the clover.  As a cluster of clover expands, it can increase soil-N levels.  Consequently, grass may dynamically exploit the clover and compete more effectively \cite{Schwinning_1996b}.  N-availability could have increased during the growing season; perhaps between-species competition decelerated the advance of the clover.

Cain et al. [1995] report that clover's stolon growth self-regulates at high density.  Sufficiently strong self-regulation among source ramets might reduce translocation to sink ramets, decelerating clover's advance.  However, biotic effects on velocity, in our experiment, were likely overwhelmed by seasonal change in temperature and day length.

Summarily, the theory of kinetic roughening offers a new, practical framework for understanding spatial pattern in both within and between-species interactions.  Furthermore, the theory's scaling predictions suggest a novel approach to linking local interface measures to properties of an invasive front at extended spatial scales.

\vspace*{0.75truecm}
\begin{center}
{\large{A}}CKNOWLEDGEMENTS
\end{center}

We thank K. Bolton, K. Shukla, and the staff at the Guelph Turfgrass Institute and Research Station for help with the experiment.  We appreciate discussion with G. Robinson, L. O'Malley, A.J. Parsons and A.C. Gorski.  This material is based upon research supported by the National Science Foundation under Grant No. DEB 0918392 (TC), DEB 0918413 (GK), and DMR 1246958 (GK); field research was supported by grants from the Ontario Ministry of Agriculture (JAN) and the Canadian Natural Sciences and Engineering Research Council (JAN).



\newpage
\begin{center}
\textbf{SUPPLEMENTAL MATERIAL}
\end{center}

\appendix

\section{Pulled versus pushed fronts}
\label{pushed_front} Models for spatial invasion often adopt a
reaction-diffusion formalism, treating population densities as
continuous variables.  A deterministic reaction-diffusion system may
yield an analytic approximation for invasion speed, given by the
asymptotic velocity of a traveling wave
\cite{Andow_1990,Caraco_2002,Murray_2003}.  But traveling waves
invoke infinitesimal population densities
\cite{Durrett_1994tpb,Pachepsky_2011}, and the linearized front can
be ``pulled'' by reproduction and dispersal of the invader at
locations where its population density is near 0
\cite{Lewis_1993,Snyder_2003}.  Deterministic reaction-diffusion
theory neglects the discreteness of individuals, the fundamental
source of endogenous, random fluctuations, and consequently
overestimates the velocity of an individual-based, dispersal-limited
dynamics \cite{Escudero_2004}.  Therefore, deterministic
reaction-diffusion equations, and their generalizations,
oversimplify the dynamics of rarity \cite{Clark_2003}; they cannot
capture consequences of strong dispersal limitation, in particular,
the spatially correlated variability along the interface we studied
experimentally.

Discrete (individual-based) models reveal effects of nonlinear,
stochastic growth processes driving an ecological interface
\cite{Wilson_1998,Moro_2001}.  Discrete models predict
front-propagation behaviors that differ from results of
deterministic diffusion models \cite{vSaarloos_2003}.  When
dispersal is limited to a local neighborhood, the interface is
``pushed'' at a velocity less than that of the corresponding
deterministic diffusion model \cite{Moro_2003}.  As a stochastic
model's interface roughens, the distance over which density
fluctuations are correlated grows \cite{Racz_1988,Majumdar_2004};
consequently, the front-runner's lead is an extreme value among
dependent random variables.

Given the assumption that a spatially clustered invader displaces a
resident competitor, the front is pushed into a meta-stable medium
\cite{Korniss_JTB2005}.  If the same invader were to propagate into
empty space (i.e., a region where the invader does not encounter
biotic resistance), the front would be pushed into an unstable
medium.  Invasion velocity is, of course, slower in the former case
\cite{Allstadt_2009}.  Interestingly, the two fronts will roughen
similarly; the same interface-length dependent scaling will emerge.

\section{Roughening and Universality}
\label{fractals} Recall the general scaling relationships of a
self-affine interface \cite{Barabasi_1995}.  During development,
roughness increases with time according to $\langle w^2(L, t)\rangle
\sim t^{2 \beta}$.  The time of crossover to statistical equilibrium
increases with interface length according to $t_{\times} \sim L^z$.
After saturation, roughness increases with interface length
according to $\langle w^2(L, \infty ) \sim L^{2 \alpha}$.
\begin{figure}[t]
  \centering
  \vspace*{3.80truecm}
  \includegraphics{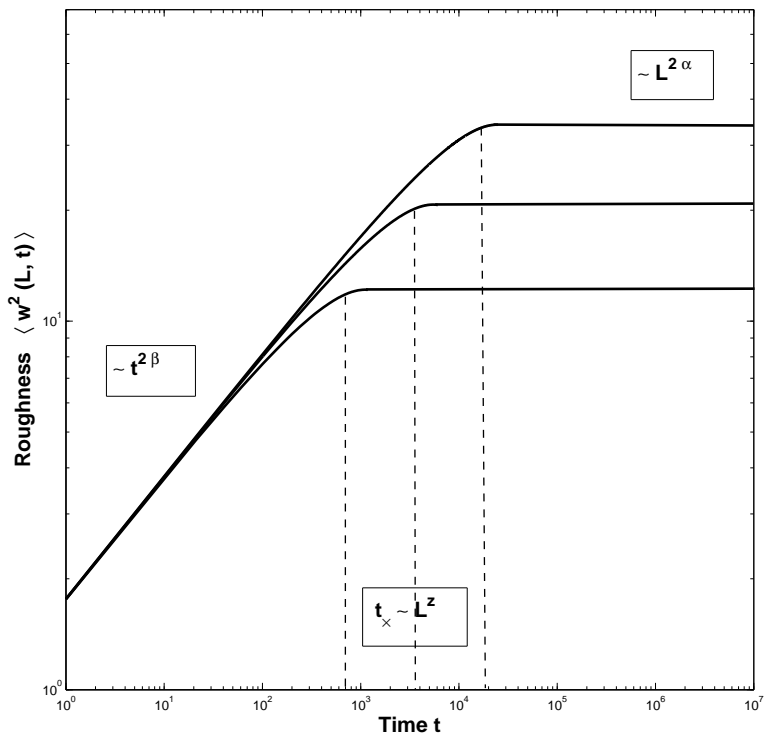}
  \vspace*{5.00truecm}
\caption{\small
Roughening over time, both scaled logarithmically.  $\langle w^2 \rangle$ increases as a power law during interface development.  Time of crossover and degree of roughening at saturation both scale with interface length.
}
\label{scaleplot}
\end{figure}
Different models for individual-level demographic processes driving
invasion may exhibit the same dependence of roughening on time, and
the equilibrium width may exhibit the same dependence on interface
length.  Such roughened interfaces belong to the same ``universality
class;'' universality offers powerful generalization.  O'Malley et
al. (2006) analyzed a model for an advancing front in a habitat
where dispersal-limited species compete for growth sites.  They
found that the model's roughening behavior belongs to the KPZ
universality class, for Kardar-Parisi-Zhang \cite{Kardar_1986}.  For
the broad class of models exhibiting KPZ universality, roughening of
a one-dimensional interface (hence the habitat has two dimensions)
implies that the dynamic exponent $z$ = 3/2, the growth exponent
$\beta$ = 1/3, and the roughening exponent $\alpha$ = 1/2.

Despite successful application of KPZ scaling relationships to a
series of real-world questions, the model's assumptions are fragile.
The nonlinear stochastic differential equation underlying the
derivation of the scaling exponents includes additive Gaussian
noise, uncorrelated in space and time.  If the noise, instead, has
spatial or temporal power-law correlation, or if the noise remains
an uncorrelated, but non-Gaussian process, then the exponents change
\cite{Sneppen_1992}.

\section{Field Methods}
\label{AndyAtGTI}

Initial monocultures were established in Fall 2007, with Dutch white
clover seed and a perennial ryegrass mix planted at respective
densities of $> 1.28$ kg/100 $m^2$, and $> 7.5$ kg/ha. For ease of
planting and establishment, plots were arranged (with one exception)
so the initial monocultures of a plot border the monoculture of the
same species in the next row.  Experimental blocks were arranged
linearly from the northeast to the southwest. Spatial constraints
required two rows within each block, aligned  from the northwest to
the southeast. One row in each block contained plots of $L$ = 1 and
16 $m$ side by side, separated by their buffer areas, plus an
additional one meter gap to ensure independence  of the plots. The
other row contained plots of the remaining $L$ in the same manner.
The order of the rows within the block and the position of plots
within a row were randomly selected.  Blocking exerted no
significant statistical effects on the results.

The ryegrass mix consisted of 40\% Barclay, 30\% Passport, and 30\%
Goalkeeper varieties.  The area was watered as necessary, and the
well drained, sandy loam soil prevented excessive moisture
accumulation.  The ryegrass required fertilization twice before it
became fully established (on 7/7/2008 and 9/19/2008; each time we
applied 25kgN/ha).  To remove weeds without disturbing the soil, we
sprayed herbicide twice (7/7/2008 and 9/19/08). The clover was
sprayed with a grass control herbicide (\emph{Poast Ultra}, 1L/ha),
and the grass with a broad-leaf control herbicide (\emph{Par 3},
55mL/100$m^2$). Throughout the experiment, we removed weeds
manually, unless removal would disturb the interface.

In spring 2009, monocultures achieved densities sufficient for the
experiment. On May 20th, plastic barriers between monocultures were
removed. Plots were mowed weekly to 5 $cm$ above ground through the
end of October 2009.  There was very little advance of the front in
this first season (no movement in most plots) possibly due to the
intense mowing regime.  In 2010, we mowed only once a month to 8
$cm$ above ground, and the clover steadily advanced.  We took
monthly photos just after mowing.  Throughout the experiment, weeds
surviving mowing were manually removed, provided their removal did
not disturb the advancing interface.

\end{document}